\documentclass[11pt]{article}
\usepackage{amsmath}
\usepackage{epsfig,epsf,fancybox}
\usepackage{amsmath}
\usepackage{amssymb}
\usepackage{graphicx}
\usepackage{color}
\usepackage{caption}
\usepackage{float}

\newtheorem{lemma}{Lemma}

\newcommand{\RR}{\mathbf R}

\begin{document}

\title{On the Continuous Fermat-Weber Problem for a Convex Polygon Using Euclidean Distance}
\author{Thomas T.C.K.\ Zhang  \thanks{Wayzata High School, MN} \and John G.\ Carlsson \thanks{University of Minnesota}}
\date{\today}
\maketitle

\begin{abstract}
In this paper, we consider the continuous Fermat-Weber problem, where the customers are continuously (uniformly) distributed along the boundary of a convex polygon. We derive the closed-form expression for finding the average distance from a given point to the continuously distributed customers along the boundary.  A Weiszfeld-type procedure is proposed for this model, which is shown to be linearly convergent. We also derive a closed-form formula to find the average distance for a given point to the entire convex polygon, assuming a uniform distribution. Since the function is smooth, convex, and explicitly given, the continuous version of the Fermat-Weber problem over a convex polygon can be solved easily by numerical algorithms.
\end{abstract}

\section{Introduction} \label{intro}

It was in the 17th century when the problem of minimizing the total distance to a certain number of set points, often known as the $1$-median problem, was first considered. Pierre de Fermat proposed a problem in which, given three points, one was to find the point at which total distance from the point to the given three points was minimum. Torricelli soon provided a geometric proof using properties of triangles, and for a while afterward, no further significant advancements occurred. More than two and a half centuries later, Alfred Weber \cite{W09} proposed an extension of Fermat's problem, in the context of minimizing transportation costs to serve a given set of customers. Therefore, the formulation is popularly known as the Fermat-Weber problem.

In 1937, Weiszfeld \cite{W37} introduced a method to solve the Fermat-Weber problem. Weiszfeld's method worked well in practice; however, a convergence analysis was missing. Quite a few papers were devoted to this topic; see e.g.~\cite{AK09,CHK05,CT89,K73}. In particular, Harold Kuhn \cite{K73}, in 1973, provided a convergence proof for the Weiszfeld method, albeit incomplete. For a modern treatment of the method and recent developments, the reader is referred to \cite{V05}, and for more information on general location theory, the reader is referred to \cite{DH04}.

In 2005, Fekete, Mitchell, and Beurer \cite{FMB05}, proposed the continuous version of the Fermat-Weber problem, in which there are infinite number of customers distributed continuously in a certain region. However, they only considered the $L_1$ norm instead of the original Euclidean norm due to complications in integration. On the topic of facility location problems in continuous space, Carlsson, Jia, and Li \cite{CJL13} discuss a further extension of the Fermat-Weber problem known as the $k$-medians problem.

In this paper, we consider the natural continuous version of the Fermat-Weber problem, in which we use the Euclidean distance. The main contribution of this paper is to present a closed-form expression for the continuous Fermat-Weber problem where the area is a polygon. Since the function is known to be smooth and convex, any existing optimization techniques for smooth convex optimization would be applicable. However, in most parts of the paper, we focus on a variant of the model, where the customers are located continuously along the edges of a convex polygon. We believe this particular model is novel and has interesting applications, for instance, finding the optimal location for a border patrol base. Due to the special structure of the problem, we are able to generalize the Weiszfeld procedure to this continuous version of the Fermat-Weber problem. The linear convergence property of the procedure is shown to hold in general.

The organization of the paper is as follows. In Section \ref{line}, we find a closed-form expression of the average distance from a given point to a given line segment and extend the expression to a convex polygon. In Section \ref{weiszfeld}, we present our variant of the Weiszfeld method to find the point in a given polygon that minimizes average distance to its boundary. In Section \ref{converge}, we prove the linear convergence of our generalized Weiszfeld procedure. Finally, in Section \ref{area}, we derive the closed-form expression of the average distance from a given point to the entire continuous area of a given polygon.

\section{Computing the Average Distance from a Point to a Line Segment} \label{line}
In order to find the average distance from a given interior point to the boundary of a polygon, we must first find a formula to compute the average distance from a point to a line segment.

Given a point and a line segment, let the point be the origin on a Cartesian plane, and one end of the line segment be $(a,0)$, and the other end be $(0,b)$. Let there be a function
$$
f(t)=(ta, (1-t)b) = \left( \begin{array}{c} 0 \\ b \end{array} \right) + t \cdot \left( \begin{array}{c} a \\ -b \end{array} \right)
$$
where $t$ is the ratio of the distance from $(0,0)$ to a point between $(0,0)$ and $(a,0)$ over length~$a$.

\begin{figure}
  \centering
  \includegraphics[width=0.4\textwidth]{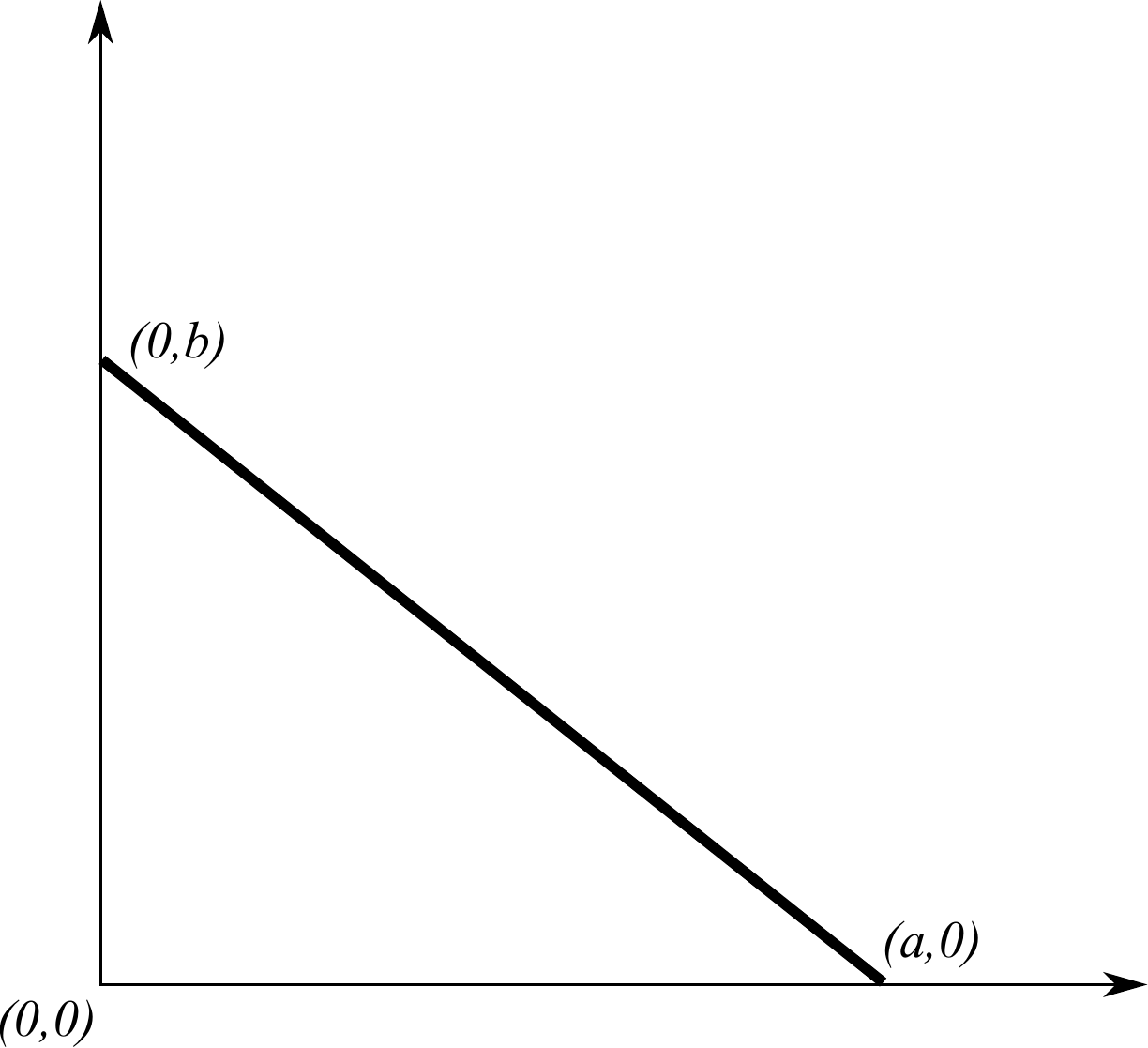}\\
  \caption{} \label{abline}
\end{figure}

The distance from $(0,0)$ to $x(t)$ can therefore be defined:
\[
\mbox{dist}(0,x(t))= \left\| \left( \begin{array}{c} 0 \\ b \end{array} \right) + t \cdot \left( \begin{array}{c} a \\ -b \end{array} \right)  \right\|=\sqrt{t^2a^2 + (1-t)^2b^2}.
\]

We can then take the integral of $\mbox{dist}(0,x(t))$
\begin{eqnarray*}
 & & \int_{0}^{1} \sqrt{t^2a^2 + (1-t)^2b^2}\: dt\\
 &=& \int_{0}^{1} \sqrt{(a^2+b^2)t^2 - 2 b^2 t +b^2}\: dt\\
 &=& \int_{0}^{1} \sqrt{(a^2+b^2) \cdot \left(t - \frac{b^2}{a^2+b^2}\right)^2 + \frac{a^2b^2}{a^2+b^2}}\: dt.
\end{eqnarray*}

\begin{lemma}\label{Lemma 1}
It holds that
$$\int \sqrt{\alpha t^2+\beta}\: dt = \frac{1}{2 \alpha} \cdot\left( \alpha \sqrt{ \alpha t^2+ \beta} \cdot t + \beta \sqrt{\alpha} \cdot \sinh^{-1}\left(\sqrt{\frac{\alpha}{\beta}}\, t \right)\right)+ C$$
where $\alpha$ and $\beta$ are positive constants, and $\sinh^{-1}$ is the inverse of the hyperbolic sine function
\[
\sinh(x)= \frac{e^x-e^{-x}}{2}.
\]
\end{lemma}

Using Lemma \ref{Lemma 1}, and treating $(a^2 + b^2)$ as $\alpha$, $\frac{a^2 b^2}{a^2 + b^2}$ as $\beta$, and $t-\frac{b^2}{a^2+b^2}$ as $t'$, we get
\begin{eqnarray*}
 & & \int_{0}^{1} \sqrt{(a^2+b^2) \cdot \left(t - \frac{b^2}{a^2+b^2}\right)^2 + \frac{a^2b^2}{a^2+b^2}}\:dt  \\
 &=& \left. \frac{1}{2a^2+2b^2} \left((a^2+b^2)\sqrt{(a^2+b^2) \cdot \left(t - \frac{b^2}{a^2+b^2}\right)^2 + \frac{a^2b^2}{a^2+b^2}}\,\cdot\, t + \frac{a^2 b^2}{\sqrt{a^2+b^2}}\cdot \sinh^{-1}\left(\frac{a^2+b^2}{ab}\cdot t \right) \right) \right|_0^1\\
 &=& \frac{1}{2a^2+2b^2} \left((a^2+b^2)\sqrt{(a^2+b^2) \cdot \left(\frac{a^2}{a^2+b^2}\right)^2 + \frac{a^2b^2}{a^2+b^2}} + \frac{a^2 b^2}{\sqrt{a^2+b^2}}\cdot \sinh^{-1}\left(\frac{a^2+b^2}{ab}\right) \right).
\end{eqnarray*}

In general, when given point $X(x,y)$ and a line segment from $P_1(a_1,b_1)$ to $P_2(a_2,b_2)$,
\begin{eqnarray*}
 & &\mbox{dist}((x,y),P(t)) \\
 &=&\sqrt{\left(x-(t \cdot a_1+(1-t) \cdot a_2)\right)^2 + \left(y-(t \cdot b_1 + (1-t) \cdot b_2 \right)^2} \\
 &=& \sqrt{(x-a_2)^2 + (y-b_2)^2 + 2\left((x-a_2)(a_2-a_1)+(y-b_2)(b_2-b_1)\right) t + \left((a_2-a_1)^2 +(b_2-b_1)^2\right) t^2}.
\end{eqnarray*}

\begin{figure}[H]
  \centering
  \includegraphics[width=0.4\textwidth]{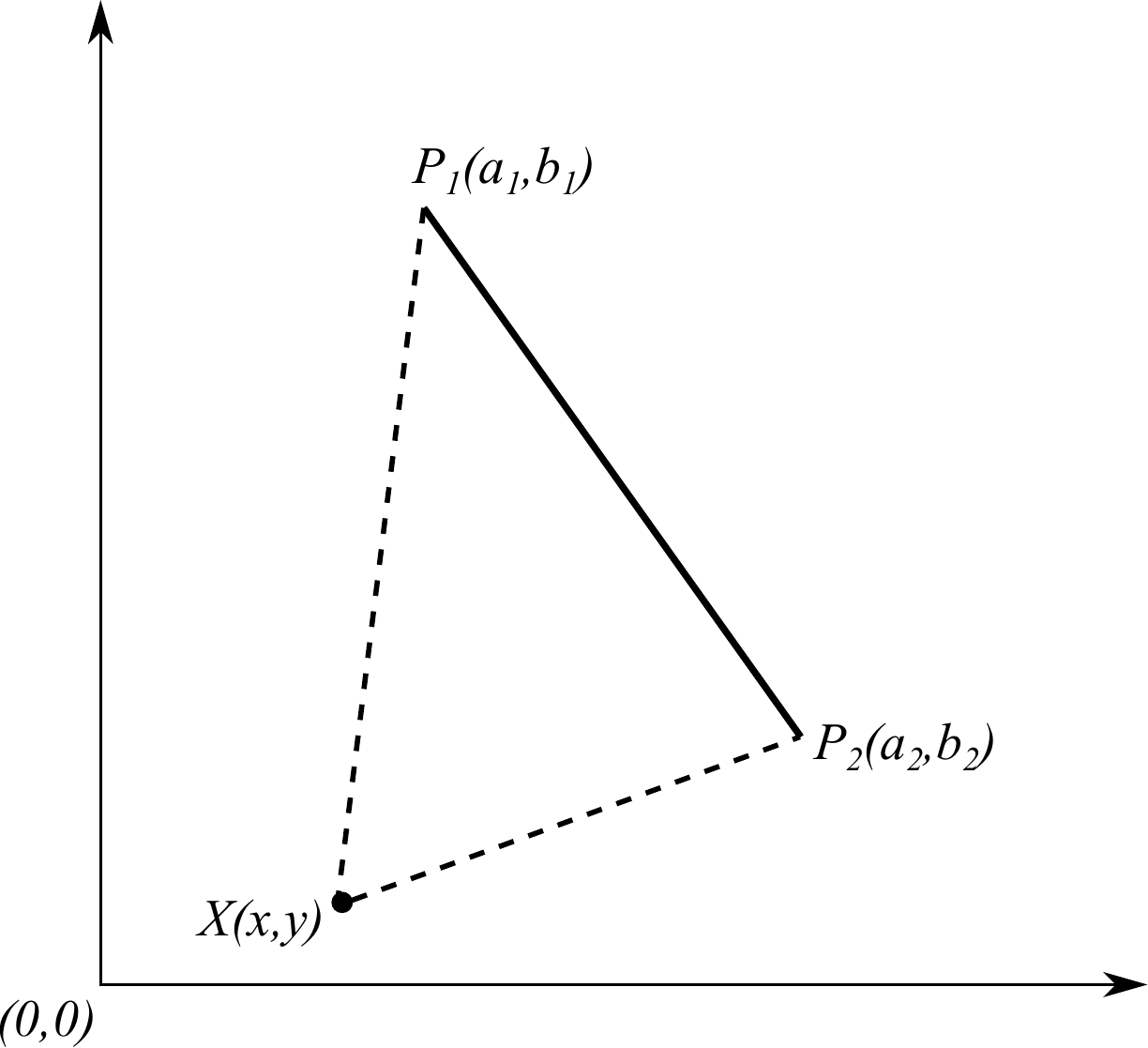}\\
  \caption{}\label{fig:p1p2line}
\end{figure}

Average distance from the point to the line segment would therefore be
\begin{eqnarray}
 & & \int_{0}^{1} \sqrt{(x-a_2)^2 + (y-b_2)^2 + 2((x-a_2)(a_2-a_1)+(y-b_2)(b_2-b_1))t + ((a_2-a_1)^2 + (b_2-b_1)^2)t^2} \:dt \nonumber \\
 &=& \int_{0}^{1} \sqrt{\|P_1-P_2\|^2 \left(t+ \frac{\langle X-P_2,P_2-P_1 \rangle}{\|P_1-P_2\|^2}\right)^2 + \frac{\|P_1-P_2\|^2 \cdot \|X-P_2\|^2 - \langle X-P_2,P_2-P_1 \rangle^2}{\|P_1-P_2\|^2}} \:dt \nonumber \\
 &=& \left.\frac{1}{2} \sqrt{\alpha t^2 + \beta} \cdot t \right|_{\gamma}^{1+\gamma} + \left. \frac{\beta}{2\sqrt{\alpha}} \sinh^{-1} \left( \sqrt{\frac{\alpha}{\beta}} \cdot t \right) \right|_{\gamma}^{1+\gamma}\nonumber \\
 &=&  \frac{1}{2} \left( 1+ \gamma \right) \sqrt{\alpha \left( 1+ \gamma \right)^2+\beta} \cdot - \frac{1}{2}\gamma \sqrt{\alpha \gamma^2+\beta}+ \frac{\beta}{2\sqrt{\alpha}} \sinh^{-1} \left( \sqrt{\frac{\alpha}{\beta}} \cdot \left( 1+ \gamma \right)\right)- \frac{\beta}{2\sqrt{\alpha}} \sinh^{-1} \left( \sqrt{\frac{\alpha}{\beta}} \gamma \right)\nonumber \\
 &=:& F(X,P_1,P_2), \label{definitionF}
\end{eqnarray}
where
\begin{eqnarray*}
\alpha &=& \|P_1-P_2\|^2 \\
\beta &=& \frac{\|P_1-P_2\|^2 \cdot \|X-P_2\|^2 - \langle X-P_2,P_2-P_1 \rangle^2}{\|P_1-P_2\|^2} \\
\gamma &=& \frac{\langle X-P_2,P_2-P_1 \rangle}{\| P_1-P_2 \|^2}.
\end{eqnarray*}



To find the average distance from a given point to the boundary of a given $n$-sided polygon $P$, with vertices at $P_1(a_1,b_1), P_2(a_2,b_2), ..., P_n(a_n,b_n)$ in clockwise order, and the given point $X(x,y)$, we treat each edge of the polygon, $\overline{P_1 P_2}, \overline{P_2 P_3}, ..., \overline{P_{n-1} P_n}, \overline{P_n P_1}$, as a line segment on which we can apply the general formula \eqref{definitionF}.
\begin{figure}
  \centering
  \includegraphics[width=0.7\textwidth]{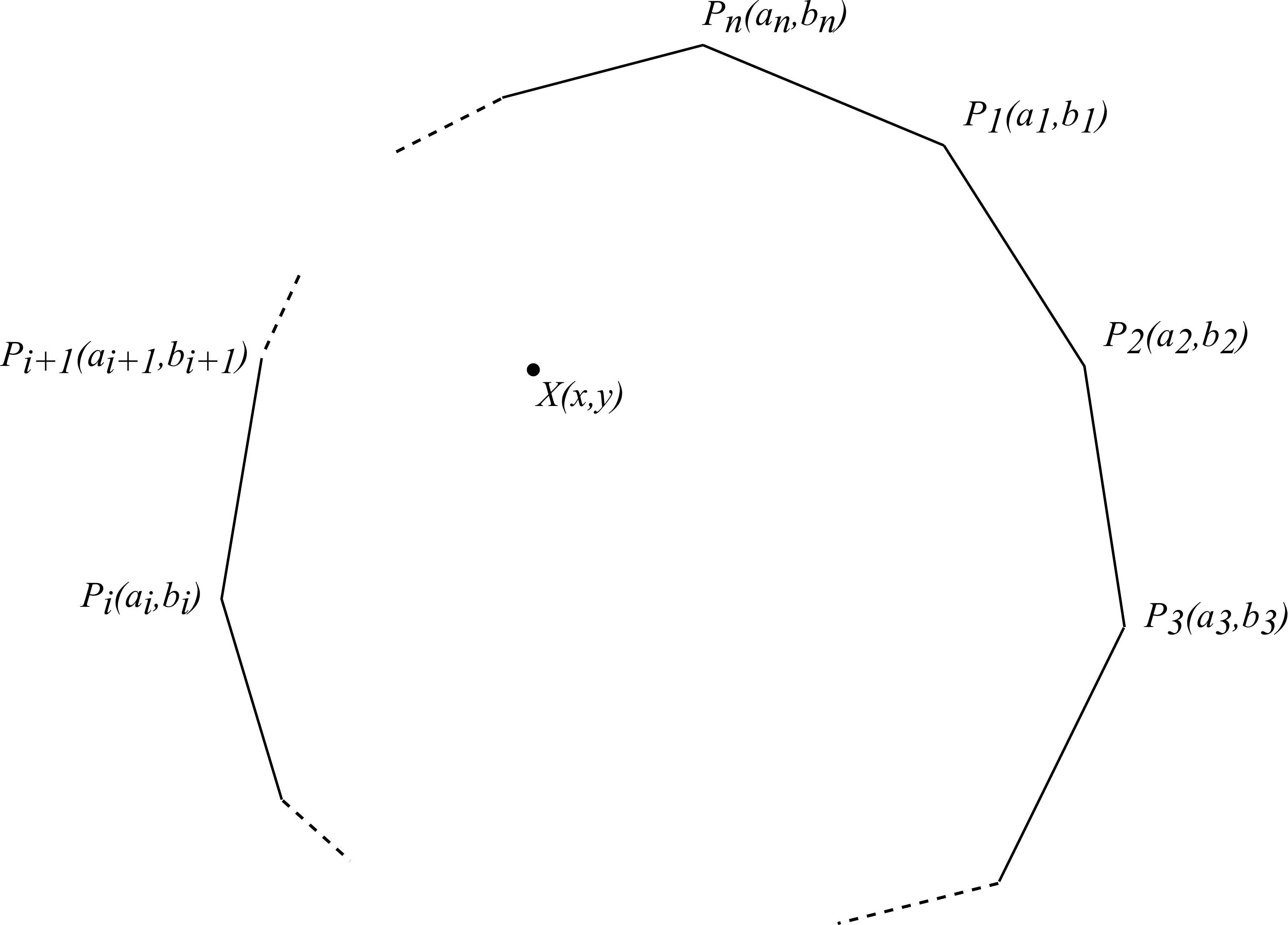}\\
  \caption{An $n$-sided polygon and point at $X(x,y)$} \label{polygon}
\end{figure}
Therefore, the average distance from the given point to all the edges is
\begin{equation} \label{distancetotheboundary}
\bar{d}(X,\partial P) = \sum_{i=1}^{n} F(X,P_i,P_{i+1})
\end{equation}
where $P_{n+1}$ is defined as $P_1$.

\section{Finding the Point Such That Average Distance from the Point to the Boundary of a Convex Polygon is Minimum} \label{weiszfeld}

In order to find the point such that average distance from the point to the boundary of a convex polygon is minimum, we see that we may treat our problem as a Weber Problem.  We can then use the Weiszfeld procedure to find the optimal point.

Let
\begin{eqnarray*}
A_i(t) &=& \left(t + \frac{\langle X-P_{i+1},P_{i+1}-P_i\rangle}{\|P_i-P_{i+1}\|^2}\right)^2 \\
B_i &=& \frac{\|P_{i+1}-P_{i}\|^2 \cdot \|X-P_{i+1}\|^2-\langle X-P_{i+1},P_{i+1}-P_{i}\rangle^2}{\|P_i-P_{i+1}\|^4}.
\end{eqnarray*}

Therefore,
\[
\nabla_x \bar{d}(X,\overline{P_iP_{i+1}}) = \|P_i-P_{i+1}\| \int_0^1 \frac{t\frac{P_{i+1}-P_{i}}{\|P_i-P_{i+1}\|^2} + \frac{X-P_{i+1}}{\|P_i-P_{i+1}\|^2}}{\sqrt{A_i(t) + B_i}} dt.
\]
To minimize the function $\bar{d}(X,\partial P)$ as defined in \eqref{distancetotheboundary}, its optimality condition yields
\begin{eqnarray*}
0 &=& \nabla_x \bar{d}(X,\partial P) \\
&=& \sum_{i=1}^{n} \nabla_x \bar{d}(X,\overline{P_iP_{i+1}}) \\
&=& -\int_0^1 \sum_{i=1}^{n} \frac{tP_i+(1-t)P_{i+1}}{\sqrt{A_i(t) + B_i}}dt + X \cdot \int_0^1 \sum_{i=1}^{n} \frac{\frac{1}{\|P_i-P_{i+1}\|}}{\sqrt{A_i(t) + B_i}}dt
\end{eqnarray*}
which can be written as
\[
X = \left(\sum_{i=1}^{n} \frac{1}{\|P_i - P_{i+1}\|} \int_0^1 \frac{tP_i+(1-t)P_{i+1}}{\sqrt{A_i(t) + B_i}}dt\right) \cdot \left(\sum_{i=1}^{n} \frac{1}{\|P_i - P_{i+1}\|} \int_0^1 \frac{dt}{\sqrt{A_i(t) + B_i}}\right)^{-1}.
\]

Let
\begin{eqnarray*}
A_i^k(t) &=& \left(t + \frac{\langle X^k-P_{i+1},P_{i+1}-P_i\rangle}{\|P_i-P_{i+1}\|^2}\right)^2 \\
B_i^k &=& \frac{\|P_{i+1}-P_{i}\|^2 \cdot \|X^k-P_{i+1}\|^2-\langle X^k-P_{i+1},P_{i+1}-P_{i}\rangle^2}{\|P_i-P_{i+1}\|^4}.
\end{eqnarray*}

The Weiszfeld procedure would be the following iterative process:
\[
X^{k+1} := \left(\sum_{i=1}^{n} \frac{1}{\|P_i - P_{i+1}\|} \int_0^1 \frac{tP_i+(1-t)P_{i+1}}{\sqrt{A_i^k(t) + B_i^k}}dt\right) \cdot \left(\sum_{i=1}^{n} \frac{1}{\|P_i - P_{i+1}\|} \int_0^1 \frac{dt}{\sqrt{A_i^k(t) + B_i^k}}\right)^{-1}.
\]

\begin{lemma}\label{Lemma 2}
Let $\alpha>0$, then
\begin{eqnarray*}
\int \frac{dt}{\sqrt{t^2+\alpha}}&=&\sinh^{-1}\left(\frac{t}{\sqrt{\alpha}}\right)+C\\
\int \frac{tdt}{\sqrt{(t+\beta)^2+\alpha}}&=&\frac{(t+\beta)^2+\alpha - \beta\sqrt{(t+\beta)^2+\alpha}\cdot \sinh^{-1}\left(\frac{t+\beta}{\sqrt{\alpha}} \right)}{\sqrt{(t+\beta)^2+\alpha}}+C.
\end{eqnarray*}
\end{lemma}
This allows us to rewrite the Weiszfeld procedure more explicitly. Let
\begin{eqnarray*}
\alpha_i^k&=&\frac{\|P_{i+1}-P_{i}\|^2 \cdot \|X^k-P_{i+1}\|^2-\langle X^k-P_{i+1},P_{i+1}-P_{i}\rangle^2}{\|P_i-P_{i+1}\|^4}\\
\beta_i^k&=&\frac{\langle X^k - P_{i+1},P_{i+1}-P_i \rangle}{\|P_i-P_{i+1}\|^2}.
\end{eqnarray*}
The above lemma gives us a closed form to express:
\begin{eqnarray*}
c_i^k &:= \int_0^1 \frac{tdt}{\sqrt{(t+\beta_i^k)^2+\alpha_i^k}} &= \sinh^{-1}\left(\frac{t}{\sqrt{\alpha_i^k}}\right)\\
d_i^k &:= \int_0^1 \frac{dt}{\sqrt{(t+\beta_i^k)^2+\alpha_i^k}} &= \frac{(t+\beta_i^k)^2+\alpha_i^k - \beta_i^k\sqrt{(t+\beta_i^k)^2+\alpha_i^k}\cdot \sinh^{-1}\left(\frac{t+\beta_i^k}{\sqrt{\alpha_i^k}} \right)}{\sqrt{(t+\beta_i^k)^2+\alpha_i^k}}.
\end{eqnarray*}

The Weiszfeld procedure is:
\begin{equation} \label{weiszfeldprocedure}
X^{k+1} = \frac{\sum_{i=1}^{n}\frac{P_{i+1}\cdot d_i^k+(P_i-P_{i+1})c_i^k}{\|P_i-P_{i+1}\|}}{\sum_{i=1}^{n} \frac{d_i^k}{\|P_i-P_{i+1}\|}}.
\end{equation}

\section{The Linear Convergence of the Weiszfeld Procedure} \label{converge}
Let $\Omega \subseteq \RR^2$ be a convex region and $S$ be its boundary, and
\begin{equation}\label{cfw}
f(X) = \int_{\xi \in S} \rho(\xi) \|X-\xi\| d\xi
\end{equation}
where $\rho$ is the density. In our above discussion, we consider the uniform distribution, where $\rho$ is a constant.

Observe that $f(X)$ is a strongly convex function for $X$ in the interior of $\Omega$. We can compute the gradient of $f(X)$ to be
\[
\nabla f(X) = \int_{\xi \in S} \rho(\xi) \frac{X-\xi}{\|X-\xi\|} d\xi,
\]
where $X$ is in the interior of $\Omega$.

The Weiszfeld procedure can be written as
\begin{eqnarray}
 X^{k+1} &=& \frac{\int_{\xi \in S} \rho(\xi) \frac{\xi}{\|X^k-\xi\|} d\xi}{\int_{\xi \in S} \rho(\xi) \frac{1}{\|X^k-\xi\|} d\xi} \nonumber\\
 &=& X^k - \frac{\nabla f(X^k)}{\int_{\xi \in S} \rho(\xi) \frac{1}{\|X^k-\xi\|} d\xi} \label{weiszfeldconvergence} \\
 &=:& T(X^k)\nonumber.
\end{eqnarray}
We want to prove that if we start from the initial point $X^0$ in the interior of $\Omega$, then the Weiszfeld procedure is descent and converges linearly to the optimum. Now, for a fixed $Y$ in the interior of $\Omega$, we introduce a quadratic function in $S$ defined as
\[
q(X;Y) = \int_{\xi \in S} \frac{\rho(\xi)}{\|\xi - Y\|} \|\xi - X\|^2 d\xi.
\]
It is easy to verify that
\begin{eqnarray*}
 q(X;X) &=& f(X), \\
 \nabla q(X;X) &=& \nabla f(X), \\
 \nabla^2 q(X;Y) &=& \left( 2\int_{\xi \in S} \frac{\rho(\xi)}{\|\xi - Y\|} d\xi \right) \cdot I.
\end{eqnarray*}
The Weiszfeld procedure, as described in \eqref{weiszfeldconvergence} can be interpreted as finding the minimum point of $q(X;Y)$, for given $Y$, namely
\[
T(Y) = \mbox{arg}\min_X q(X;Y).
\]
Using the above facts, on one hand, we have
\begin{eqnarray}
q(Y;Y) - q(T(Y);Y) &\ge& 2\int_{\xi \in S} \frac{\rho(\xi)}{\|\xi - Y\|} d\xi \cdot \|T(Y) - Y\|^2 \nonumber \\
&=& \frac{2}{\int_{\xi \in S} \frac{\rho(\xi)}{\|\xi - Y\|} d\xi} \|\nabla f(Y)\|^2. \label{secondorderterm}
\end{eqnarray}
On the other hand,
\begin{eqnarray}
& & q(T(Y);Y) \nonumber \\
&=& \int_{\xi \in S} \frac{\rho(\xi)}{\|\xi - Y\|} \|\xi - T(Y)\|^2 d\xi \nonumber \\
&=& \int_{\xi \in S} \frac{\rho(\xi)}{\|\xi - Y\|} \left( \|\xi - T(Y)\| - \|\xi - Y\| + \|\xi - Y\| \right)^2 d\xi \nonumber \\
&=& \int_{\xi \in S} \frac{\rho(\xi)}{\|\xi - Y\|} \left[ (\|\xi - T(Y)\| - \|\xi - Y\|)^2 + 2\|\xi - Y\|(\|\xi - T(Y)\| - \|\xi - Y\|) + \|\xi - Y\|^2 \right] d\xi \nonumber \\
&>& 2f(T(Y))-f(Y). \label{perfectsquare}
\end{eqnarray}
Since $q(T(Y);Y)<q(Y;Y) = f(Y)$, from \eqref{perfectsquare}, we have $f(Y)<f(T(Y))$. In other words, the Weiszfeld procedure improves the objective function monotonically. Let $\hat{f} = \min_{X \in \partial P} f(x)$. If we start from a point $X^0$ in the interior of $P$ such that $f(X^0)<\hat{f}$, then the iterates produced by the Weiszfeld procedure will remain in the interior of $P$. Therefore, the Weiszfeld procedure is well-defined. Moreover, combining \eqref{secondorderterm} and \eqref{perfectsquare} we have
\begin{eqnarray*}
f(Y) - f(T(Y)) &\ge& \frac{1}{\int_{\xi \in S} \frac{\rho(\xi)}{\|\xi - Y\|} d\xi} \|\nabla f(Y)\|^2 \\
&\ge& \omega \left(f(Y) - f(X^*)\right),
\end{eqnarray*}
where $X^*$ is the minimum point of $f$.  The constant $0<\omega<1$ is dependent on the level set of the initial point $X^0$, which is assumed to be in the interior of $\Omega$.  This yields
\[
f(T(Y)) - f(X^*) \le (1-\omega)(f(Y) - f(X^*)),
\]
which is the desired linear rate of convergence for the Weiszfeld procedure applied to the above function as defined in \eqref{cfw}.

\section{A Closed-form Expression for the Continuous Fermat-Weber Objective Function} \label{area}

Recall that the average distance from a given point in the polygon to one of the edges of the polygon is defined $F(X,P_i,P_{i+1})$. If we shrink the polygon with $X$ as the homothetic center, we may write the the expression $F(X,\lambda P_i + (1-\lambda)X,\lambda P_{i+1} + (1-\lambda)X)$ to express the average distance from the given point to the scaled line segment, where $\lambda$ is the scale factor. We observe that:
\[
F(X,\lambda P_i + (1-\lambda)X,\lambda P_{i+1} + (1-\lambda)X) = \lambda F(X,P_i,P_{i+1})
\]
and the perpendicular distance from $X$ to the edge $\overline{P_iP_{i+1}}$ is
\begin{equation} \label{determinant}
\mbox{dist}(X,\overline{P_iP_{i+1}}) = \det \left[ P_{i+1}-X ,\, P_i-X  \right].
\end{equation}

\begin{figure}[H]
  \centering
  \includegraphics[width=0.35\textwidth]{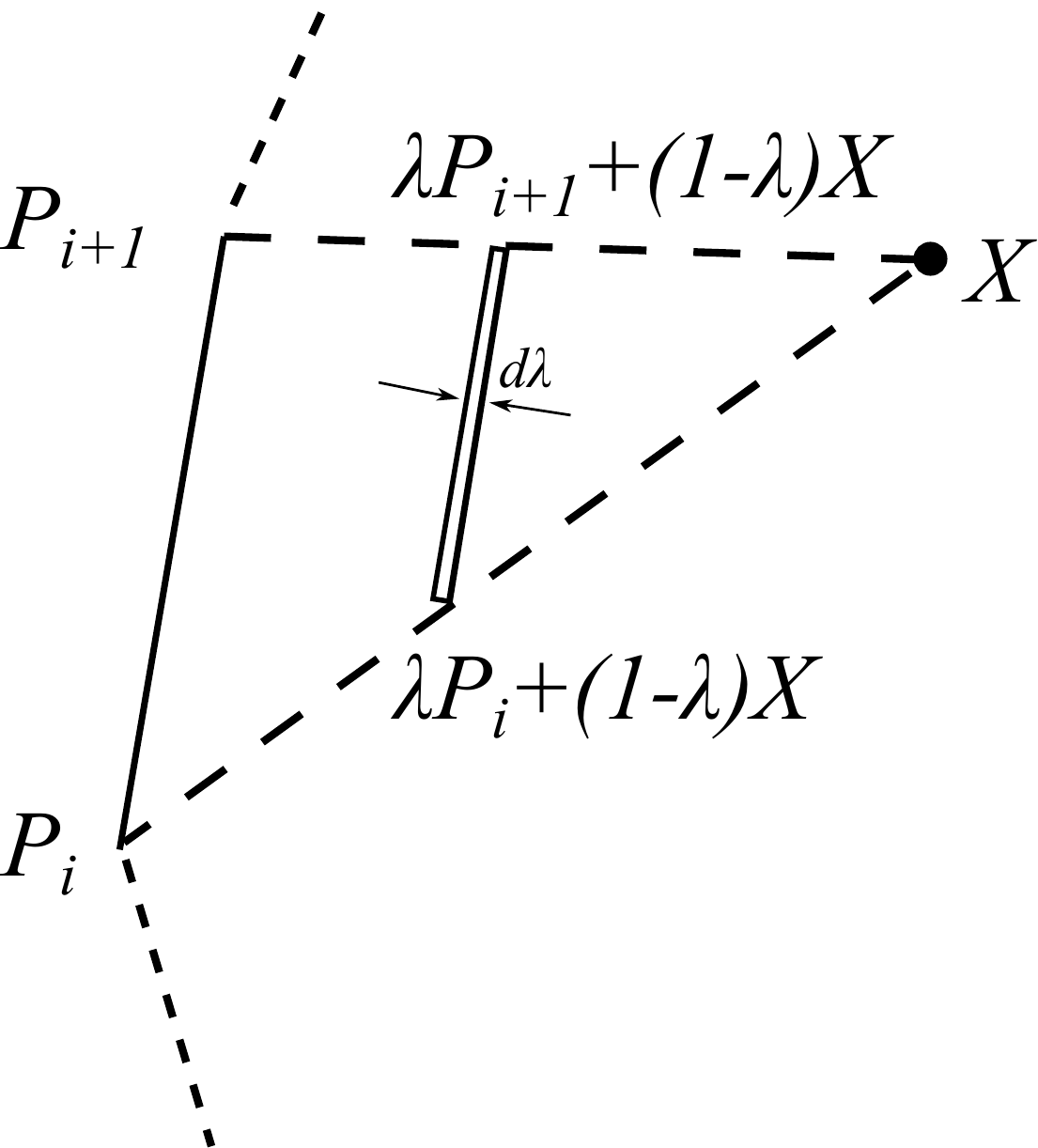}\\
  \caption{}\label{fig:polygon2}
\end{figure}

Therefore, by \eqref{determinant}, the total Fermat-Weber objective is:
\begin{eqnarray*}
 & & \sum_{i=1}^n \int_0^1 F(X,\lambda P_i + (1-\lambda)X,\lambda P_{i+1} + (1-\lambda)X) \cdot \det \left[ P_{i+1}-X ,\, P_i-X  \right] \: d\lambda \\
 &=& \sum_{i=1}^n \int_0^1 \lambda F(X,P_i,P_{i+1}) \cdot \det \left[ P_{i+1}-X ,\, P_i-X  \right] \: d\lambda \\
 &=& \frac{1}{2} \sum_{i=1}^n F(X,P_i,P_{i+1}) \cdot \det \left[ P_{i+1}-X ,\, P_i-X  \right].
\end{eqnarray*}

This is a smooth convex function from its definition; hence, it is easy to minimize using convex optimization methods.

\section{Discussions}

The implication of the last section is that the continuous Fermat-Weber problem, or the $1$-median problem, can be solved easily using the Euclidean distance. This method can be applied to solve the more complicated $k$-medians problem, at least heuristically. For instance, one heuristic solution would be the following. First, we select initial locations, then draw a Voronoi diagram based on the $k$ facilities. Since each sub-region of the Voronoi diagram is a convex polygon, we may apply the expression discussed in Section \ref{area} to each sub-region, and find the optimal position in each. We then redraw the Voronoi diagram based on the newly computed locations of the $k$ facilities. The process can be repeated iteratively until a satisfactory solution is found.

In principle, it is also possible to obtain a closed-form expression from a given point to a convex polytope in a Euclidean space with fixed dimensions. The idea is to apply induction on the dimension and to use the proportionality of the distance from the given point to the parallely positioned line segments. An example of such an extension can also be seen in Figure \ref{fig:polygon2}, but the concept can be extended to any dimension. However, the formula will likely become too complicated to be practical.






\end{document}